\magnification \magstep1
\baselineskip 18pt
\centerline{\bf On the relation between a zero-point-field-induced inertial effect}
\centerline{\bf and the Einstein-de Broglie formula}
\bigskip
\centerline{Bernard Haisch$^*$}
\centerline{\it Solar and Astrophysics Laboratory, Dept. H1-12, Bldg.
252,  Lockheed Martin}
\centerline{\it 3251 Hanover Street, Palo Alto, California 94304}
\centerline{haisch@starspot.com}
\bigskip
\centerline{Alfonso Rueda}
\centerline{\it Department of Electrical Engineering \& Department of Physics, ECS
Building}
\centerline{\it California State University, 1250 Bellflower Blvd.,
Long Beach, California 90840}
\centerline{arueda@csulb.edu}

\centerline{({\it Physics Letters A}, Vol. 268, pp. 224--227, 2000)}
\medskip
\hrule\medskip
\noindent{\bf Abstract}

It has been proposed that the scattering of electromagnetic
zero-point radiation by accelerating objects results in a reaction force that may
account, at least in part, for inertia [1,2,3]. This arises because of asymmetries in the
electromagnetic zero-point field (ZPF) or electromagnetic quantum vacuum as perceived from
an accelerating reference frame. In such a frame, the Poynting vector and momentum flux of
the ZPF become non-zero. If one assumes that scattering of the ZPF radiation takes place
at the level of quarks and electrons constituting matter, then it is possible for both
Newton's equation of motion,
${\bf f}=m{\bf a}$, and its relativistic covariant generalization, ${\cal F}=d{\cal
P}/d\tau$, to be obtained as a consequence of the non-zero ZPF
momentum flux. We now conjecture that this scattering must take place at the Compton
frequency of a particle, and that this interpretation of mass leads directly to the de
Broglie relation characterizing the wave nature of that particle in motion,
$\lambda_B=h/p$. This suggests a perspective on a connection between
electrodynamics and the quantum wave nature of matter. Attempts to extend this perspective
to other aspects of the vacuum are left for future consideration.
\medskip\hrule\medskip

\noindent $^*$Current Address: California Institute for Physics and Astrophysics, 366
Cambridge Ave., Palo Alto, CA 94306 (http://www.calphysics.org)

Using the techniques of stochastic electrodynamics we examined the Poynting vector of the
electromagnetic ZPF of the quantum vacuum in accelerating reference frames
[1,2]. This led to a surprisingly simple and intuitive relation between what should be the
inertial mass,
$m_i$,  of an object of proper volume $V_0$ and the energy density of the ZPF
instantaneously contained in $V_0$. Besides simplicity, this new approach [1,2] improved
over a previous one [3] in that it yielded a covariant generalization. As derived from
the force associated with the ZPF momentum flux in transit through the object, $m_i$ and
$\rho_{ZP}$ were found to be related as follows: 

$$m_i={V_0 \over c^2} \int \eta(\omega) \rho_{ZP}(\omega) \ d\omega , \eqno(1)
$$

\smallskip\noindent
where $\rho_{ZP}$ is the well known spectral energy density of the ZPF

$$
\rho_{ZP}(\omega) ={\hbar \omega^3 \over 2 \pi^2 c^3} . \eqno(2)
$$

\smallskip\noindent
Viewed this way, inertial mass, $m_i$,  appeared to be a peculiar form of coupling
parameter between the electromagnetic ZPF and the electromagnetically
interacting fundamental particles (quarks and electrons) constituting matter. The key to
deriving an  equation of motion (${\cal F}=d{\cal P}/d\tau$ in the relativistic case) from
electrodynamics is to assume that some form of scattering of the non-zero (in accelerating
frames) ZPF momentum flux takes place. The reaction force resulting from such scattering
would appear to be the physical origin of inertia. The parameter
$\eta(\omega)$ in eqn. (1) parametrizes such a scattering efficiency whose strength and
frequency dependence have been unknown. 

It was proposed by de Broglie that an elementary particle is associated with a localized
wave whose frequency is the Compton frequency, yielding the Einstein-de Broglie equation:
$$\hbar \omega_C=m_0 c^2 . \eqno(3)
$$
As summarized by Hunter [4]: ``\dots what we regard as the (inertial) mass of the particle
is, according to de Broglie's proposal, simply the vibrational energy (divided by $c^2$)
of a localized oscillating field (most likely the electromagnetic field). From this
standpoint inertial mass is not an elementary property of a particle, but rather a
property derived from the localized oscillation of the (electromagnetic) field. De Broglie
described this equivalence between mass and the energy of oscillational motion\dots as
{\it `une grande loi de la Nature'} (a great law of nature).'' The rest mass $m_0$ is
simply $m_i$ in its rest frame. What de Broglie was proposing is that the left-hand
side of eqn. (3) corresponds to physical reality; the right-hand side is in a sense
bookkeeping, defining the concept of rest mass.

This perspective is
consistent with the proposition that inertial mass, $m_i$, is also not a fundamental
entity, but rather a coupling parameter between electromagnetically interacting particles
and the ZPF. De Broglie assumed that his
wave at the Compton frequency originates in the particle itself. An alternative
interpretation is that a particle  ``is tuned to a wave originating in the high-frequency
modes of the zero-point background field.''[5, 6] The de Broglie oscillation would thus be
due to a resonant interaction with the ZPF, presumably the same resonance that is
responsible for creating a contribution to inertial mass as in eqn. (1). In other words,
the ZPF would be driving this $\omega_C$ oscillation.

We therefore suggest that an elementary charge driven to oscillate at the Compton
frequency, $\omega_C$,  by the ZPF may be the physical basis of the $\eta(\omega)$
scattering parameter in eqn. (1).  For the case of the electron, this would imply that
$\eta(\omega)$ is a sharply-peaked resonance at the frequency, expressed in terms of
energy, $\hbar
\omega=512$ keV. The inertial mass of the electron would physically be the reaction force
due to resonance scattering of the ZPF at that frequency.

This leads to a surprising corollary. It can be shown that
as viewed from a laboratory frame, the standing wave at the Compton frequency in the
electron frame transforms into a traveling wave having the de Broglie wavelength,
$\lambda_B=h/p$, for a moving electron. The wave nature of the moving electron appears to
be basically due to Doppler shifts associated with its Einstein-de Broglie resonance
frequency. This has been shown in detail in the monograph of de la Pe\~na and Cetto [5]
(see also Kracklauer [6]).

Assume an electron is moving with velocity $v$ in the $+x$-direction. For simplicity
consider only the components of the ZPF in the $\pm x$ directions. The ZPF-wave
responsible for driving the resonant oscillation impinging on the electron from the front
will be the ZPF-wave seen in the laboratory frame to have frequency $\omega_-=\gamma
\omega_C (1 - v/c)$, i.e. it is the wave below the Compton frequency in the laboratory
that for the electron is Doppler shifted up to the
$\omega_C$ resonance. Similarly the ZPF-wave responsible for driving the electron resonant
oscillation impinging on the electron from the rear will have a laboratory frequency
$\omega_+=\gamma \omega_C (1 + v/c)$ which is Doppler shifted down to $\omega_C$ for the
electron. The same transformations apply to the wave numbers,
$k_+$ and $k_-$. The Lorentz invariance of the ZPF spectrum ensures that regardless of the
electron's (unaccelerated) motion the up- and down-shifting of the laboratory-frame
ZPF will always yield a standing wave in the electron's frame.

It has been proposed by de la Pe\~na and Cetto [5] and by Kracklauer [6] that in the
laboratory frame the superposition of these two waves results in an apparent traveling
wave  whose wavelength is
$$\lambda = {c \lambda_C \over \gamma v} \ , \eqno(4)$$
\noindent
which is simply the de Broglie wavelength, $\lambda_B=h/p$, for a particle of momentum
$p=m_0
\gamma v$.
This is evident from looking at the summation of two oppositely moving wave trains of
equal amplitude,
$\phi_+$ and $\phi_-$, in the particle and laboratory frames [5]. In the rest frame of the
particle the two wave trains combine to yield a single standing wave.

In the laboratory frame we have for the sum,

$$\phi=\phi_+ + \phi_- =
\cos( \omega_+ t - k_+ x + \theta_+) + \cos( \omega_- t + k_- x + \theta_-) \eqno(5)
$$
where
$$\eqalignno{
\omega_{\pm}&=\omega_z \pm \omega_B &(6a) \cr
k_{\pm}&=k_z \pm k_B &(6b) }
$$
and
$$\eqalignno{
\omega_z=\gamma \omega_C \ , \ \ \ &\omega_B=\gamma \beta \omega_C &(7a) \cr
k_z=\gamma k_C \ , \ \ \ &k_B=\gamma \beta k_C \ . &(7b)  }
$$
The respective random phases associated with each one
of these independent ZPF wavetrains are $\theta_{+,-}$. After some algebra one obtains
that the oppositely moving wavetrains appear in the form

$$
\phi=2 \cos(\omega_z t - k_B x + \theta_1) \cos(\omega_B t - k_z x + \theta_2) \eqno(8)
$$
where $\theta_{1,2}$ are again two independent random phases
$\theta_{1,2}={1 \over 2}(\theta_+ \pm \theta_-)$. Observe that for fixed $x$, the rapidly
oscillating ``carrier'' of frequency $\omega_z$ is modulated by the slowly varying envelope
function in frequency $\omega_B$. And {\it vice versa} observe that at a given $t$ the
``carrier'' in space appears to have a relatively large wave number $k_z$ which is
modulated by the envelope of much smaller wave number $k_B$. Hence
both timewise at a fixed point in space and spacewise at a given time, there appears a
carrier that is modulated by a much broader wave of dimension corresponding to the de
Broglie time $t_B=2\pi/\omega_B$, or equivalently, the de Broglie wavelength
$\lambda_B=2\pi/k_B$.

This result may be generalized to include ZPF radiation from all other directions, as may
be found in the monograph of de la Pe\~na and Cetto [5]. They conclude by stating:
``The foregoing discussion assigns a physical meaning to de Broglie's wave: it is the {\it
modulation} of the wave formed by the Lorentz-transformed, Doppler-shifted superposition
of the whole set of random stationary electromagnetic waves of frequency
$\omega_C$ with which the electron interacts selectively.''

Another way of looking at the spatial modulation is in terms of the
wave function. Since

$$
{\omega_C \gamma v \over c^2} = {m_0 \gamma v \over \hbar} = {p \over \hbar}
\eqno(9)
$$ 

\smallskip\noindent
this spatial modulation is exactly the $e^{i p x / \hbar}$ wave function of a freely
moving particle satisfying the Schr\"odinger equation.
The same argument has been made by Hunter [4].
In such a view the quantum wave function
of a moving free particle becomes a ``beat frequency'' produced by the relative motion of
the observer with respect to the particle and its oscillating charge.

It thus appears that a simple model of a particle as a ZPF-driven oscillating charge
with a resonance at its Compton frequency may simultaneously offer insight into the nature
of inertial mass, i.e. into rest inertial mass and its relativistic extension, the
Einstein-de Broglie formula and into its associated wave function involving the de Broglie
wavelength of a moving particle. If the de Broglie oscillation is indeed driven by the ZPF,
then it is a form of Schr\"odinger's {\it Zitterbewegung}. Moreover there is a substantial
literature attempting to associate spin with {\it Zitterbewegung} tracing back to the work
of Schr\"odinger [7]; see for example Huang [8] and Barut and Zanghi [9]. In the context of
ascribing the {\it Zitterbewegung} to the fluctuations produced by the ZPF,
it has been proposed that spin may be traced back to the (circular) polarization of the
electromagnetic field, i.e. particle spin may derive from the spin of photons in the
electromagnetic quantum vacuum [5]. It is well known, in ordinary quantum theory, that the
introduction of $\hbar$ into the ZPF energy density spectrum $\rho_{ZP}(\omega)$ of
eqn. (2) is made via the harmonic-oscillators-quantization of the electromagnetic
modes and that this introduction of $\hbar$ is totally independent from the
simultaneous introduction of $\hbar$ into the particle spin. The idea expounded
herein points however towards a connection between the $\hbar$ in $\rho_{ZP}(\omega)$
and the $\hbar$ in the spin of the electron. In spite of a suggestive preliminary
proposal, an exact detailed model of this connection remains to be developed [10].
Finally, although we amply acknowledge [1,2] that other vacuum fields besides the
electromagnetic do contribute to inertia, e.g. see [11], no attempt has been made within
the context of the present work to explore that extension.

A standard procedure of conventional quantum physics is to limit the describable elements
of reality to be directly measurable, i.e. the so-called physical
{\it observables}. We apply this philosophy here by pointing
out that inertial mass itself does not qualify as an observable. Notions such as
acceleration, force, energy and electromagnetic fields constitute proper observables;
inertial mass does not. We propose that the inertial mass parameter can be accounted for in
terms of the forces and energies associated with the electrodynamics of the ZPF.

\bigskip\noindent
{\bf Acknowledgement}

\bigskip\noindent
This work is supported by NASA research grant NASW-5050.

\bigskip\noindent
{\bf References}

\bigskip\noindent
[1] A. Rueda and B. Haisch,
%Inertia as reaction of the vacuum to accelerated motion.
{\it Phys. Lett. A} {\bf 240}, 115 (1998).

\bigskip\noindent
[2] A. Rueda and B. Haisch,
% Contribution to inertial mass by reaction of the vacuum to accelerated motion.
{\it Foundation of Phys.} {\bf 28}, 1057 (1998).

\bigskip\noindent
[3] B. Haisch, A. Rueda, and H. E. Puthoff, H. E.,
%Inertia as a zero-point-field Lorentz force.
{\it Phys. Rev. A} {\bf 48}, 678 (1994).

\bigskip\noindent
[4] G. Hunter, 
%Electrons and photons as soliton waves,
in {\it The Present Status of the Quantum Theory of Light}, S. Jeffers et al. (eds.),
(Kluwer), pp. 37--44 (1997).

\bigskip\noindent
[5] L. de la Pe\~na and M. Cetto, {\it The Quantum Dice: An Introduction to Stochastic
Electrodynamics}, (Kluwer Acad. Publ.), chap. 12 (1996).

\bigskip\noindent
[6] A. F. Kracklauer, 
% An intuitive paradigm for quantum mechanics.
{\it Physics Essays} {\bf 5}, 226 (1992).
%also preprint ``Pilot Wave Steerage'' (1999).

\bigskip\noindent
[7] E. Schr\"odinger,
%\"Uber die kraftfreie Bewegung in der relativischen Quantenmechanik.
{\it Sitzungsbericht preuss. Akad. Wiss., Phys. Math. Kl.} {\bf 24}, 418 (1930).

\bigskip\noindent
[8] K. Huang,
%On the zitterbewegung of the Dirac electron.
{\it Am. J. Physics} {\bf 20}, 479 (1952).

\bigskip\noindent
[9] A. O. Barut and N. Zanghi,
%Classical model of the Dirac electron.
{\it Phys. Rev. Lett.}, {\bf 52}, 2009 (1984).

\bigskip\noindent
[10] A. Rueda,
%Stochastic Electrodynamics with Particle Structure: Part I: zero-point
%induced Brownian motion.
{\it Foundations of Phys. Letts.} {\bf 6}, No. 1, 75 (1993);
%Stochastic Electrodynamics with Particle Structure: Part II: Towards a zero-point
%induced wave beahviour
{\bf 6}, No. 2, 139 (1993).

\bigskip\noindent
[11] J.-P. Vigier, {\it Foundations of Phys.} {\bf 25}, No. 10, 1461 (1995).

\bye